\documentclass[sigconf, 10pt]{acmart}

\AtBeginDocument{%
  \providecommand\BibTeX{{%
    \normalfont B\kern-0.5em{\scshape i\kern-0.25em b}\kern-0.8em\TeX}}}
    
\usepackage[utf8]{inputenc}
\usepackage{hyperref}
\usepackage{caption}
\usepackage{subfigure}

\setcopyright{none}
\renewcommand\footnotetextcopyrightpermission[1]{}
\acmConference[SNCNW 2022]{17th Swedish National Computer Networking Workshop (SNCNW 2022)}{June 22}{Stockholm, Sweden.}

\usepackage{graphicx}
\usepackage{url}
\begin{document}
\title{Evolving 5G: ANIARA, an Edge-Cloud perspective}
\author{Ian Marsh}
\affiliation{%
  \institution{RISE AB, Sweden}
}

\author{Wolfgang John}
\affiliation{%
  \institution{Ericsson AB, Sweden}
}

\author{Ali Balador}
\affiliation{%
  \institution{Ericsson AB, Sweden}
}

\author{Federico Tonini}
\affiliation{%
  \institution{Chalmers University, Sweden}
}

\author{Jalil Taghia}
\affiliation{%
  \institution{Ericsson AB, Sweden}
}

\author{Andreas Johnsson}
\affiliation{%
  \institution{Ericsson AB, Sweden}
}

\author{Paolo Monti}
\affiliation{%
  \institution{Chalmers University, Sweden}
}

\author{Jonas Gustafsson}
\affiliation{%
  \institution{RISE AB, Sweden}
}

\author{Pontus Sköldström}
\thanks{Pontus Sköldström was with Ericsson AB at the time of writing this paper}
\affiliation{
  \institution{Qamcom AB, Sweden}
}



\author{Johan Sjöberg}
\affiliation{%
  \institution{Ericsson AB, Sweden}
}

\author{Jim Dowling}
 \orcid{1234-5678-9012}
\affiliation{%
  \institution{Hopsworks AB, Sweden}
}









%
%
\renewcommand{\shortauthors}{Marsh, et al.}

\ccsdesc[500]{Hardware~Energy metering}
\ccsdesc[500]{Computing methodologies~Multi-agent systems}
\ccsdesc[500]{Computing methodologies~Self-organization}

%
%
\keywords{Edge computing, federated learning, container technologies and orchestration, energy metering.}

%
%
%

\begin{abstract}

Emerging use-cases like smart manufacturing and smart cities pose challenges in terms of latency, which cannot be satisfied by traditional centralized networks. Edge networks, which bring computational capacity closer to the users/clients, are a promising solution for supporting these critical low latency services. Different from traditional centralized networks, the edge is distributed by nature and is usually equipped with limited connectivity and compute capacity. This creates a complex network to handle, subject to failures of different natures, that requires novel solutions to work in practice.
To reduce complexity, more lightweight solutions are needed for containerization as well as smart monitoring strategies with reduced overhead. Orchestration strategies should provide reliable resource slicing with limited resources, and intelligent scaling while preserving data privacy in a distributed fashion. Power management is also critical, as providing and managing a large amount of power at the edge is unprecedented.

\end{abstract}

\maketitle
%
%
\section{Introduction}

\begin{figure}[!ht]
 \centering
 \includegraphics[scale=0.42]{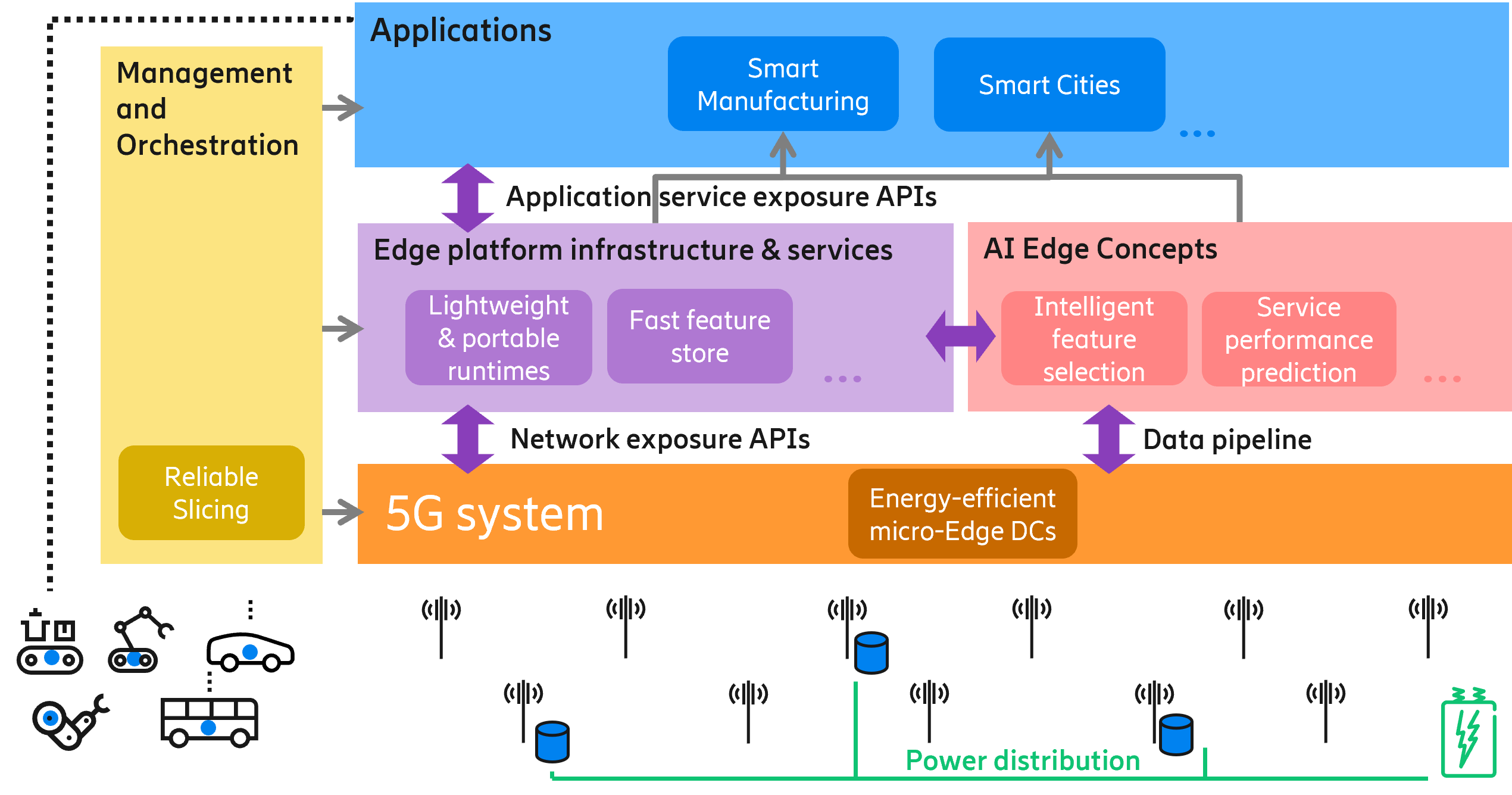}
 \caption{Functional structure of the ANIARA project.}
 \label{aniara_fig}
\end{figure}

The project is based on two use case families addressing smart manufacturing and smart cities. Figure~\ref{aniara_fig} illustrates the functional technologies for evolving 5G edge systems. 
%
%
%
%
The application scenarios are drawn upon the synergies between cloud computing, 5G, and mobile edge compute. 
Our studies include elaborating detailed use case descriptions for verticals using 5G and the edge cloud. 
ANIARA use cases cover two main aspects of factory operations, akin to \cite{DagstuhlRoadmap2021}. These are: environmental monitoring and control of the factory floor such as air quality; temperature and power management and operations monitoring and control such as robot cell control, logistics and safety. These high-level use cases drive the system architecture requirements and serve as the basis for technology-specific use cases such as private 5G, edge micro datacenters and AI at the edge. 


The remainder of this paper is organized to provide additional details of the components in Figure \ref{aniara_fig}. Section 2 provides details about edge platform infrastructure and services that have been developing in ANIARA, including lightweight and portable execution environments and fast, dependable feature stores.
Apart from edge infrastructure, developing edge AI enablers is also another important objective of the ANIARA. Section 3 describes AI/ML methods designed and developed within the project so far, including ML models for life-cycle management, intelligent feature selection and distributed learning in the edge considering privacy. ANIARA also addresses management and orchestration including both edge and cloud scenarios, to satisfy the specific needs of the use cases, detailed in Section 4. Section 5 presents research on smart power for building a large-scale 5G edge system, shown in the lower portion of Figure \ref{aniara_fig}. Section 6 provides a comprehensive related work. Finally, Section 7 gives an overview of works planned to be done in the future.

%
%
\section{Edge platform components} 
\label{section_WP3}


%
%
\subsection{Programmable, light containerisation} 
\label{WA}
\hyphenation{Web-Assembly}

Lightweight and portable execution environments have been identified as a crucial enabler for higher  flexibility and dynamism of application deployment in a distributed network \cite{wikstrom2020ever}. In ANIARA, we experiment with WebAssembly technologies to fill this gap.
WebAssembly is an \textsl{open} binary instruction format for a stack-based virtual machine, designed to support  existing programming languages in a web browser environment. In the browser, WebAssembly is generally faster than JavaScript due to its more compact format and manual memory management. The virtual machine is, however, not bound to browsers, but can be used standalone on various platforms. A deployment flow is where a high-level language such as C, C++, Rust, Python and so on can be compiled to WebAssembly (using clang) and executed on various hardware or software platforms. Webassembly allows edge applications to execute within a wide range of devices and operating systems. To prove the versatility and low footprint, we wrote an application in Rust and executed it on a WebAssembly runtime using a \$30 microcontroller. The same application was also written in Python that runs (on WebAssembly again) in an ASCII terminal.
%
%
%


%
%
%
%
%
%
%
\subsection{Fast, dependable, feature store}
\label{FS}

A production-grade AI solution for edge computing, known as Feature stores, is a part of ANIARA. They act as a halfway house between data scientists and data engineers, enabling the same feature computation code to be used for model training and inference. Feature stores also act as a centralized repository of AI models\footnote{\url{https://www.featurestore.org}} and have been adapted to perform in distributed scenarios for real-time data synchronization between geographically distributed Edge locations.
%
%
The underlying database is called RonDB, evaluated using \textbf{LATS}, meaning low \textbf{L}atency, high \textbf{A}vailability, high \textbf{T}hroughput, scalable \textbf{S}torage principles to list RonDB's performance: 
\vspace{1mm}
\begin{itemize}
   \item RonDB x3 times lower latency  
    \item RonDB 1.1M reads/sec, REDIS 800k reads/sec 
    \item RonDBs processed 200M reads in a 30 node cluster
\end{itemize}
%

%
%
\section{AI for Edge Management} 
\label{section_WP2}

We divide AI for edge management into three main research tracks: (i) ML model life-cycle management, (ii) Intelligent feature selection, and (iii) Distributed learning in the edge. The three tracks have been studied within a use case for service performance prediction from edge statistics available to an operator with the objective to automate parts of the edge management process.



%
%
\paragraph{ML model life-cycle management} Ensuring high-performing ML models that are continuously updated and correctly deployed is critical for successful integration of ML in the edge. Transfer learning, which is one approach, allows for structured incorporation of previously acquired knowledge enabling timely and robust model adaptation, especially when data are scarce for reliable training of ML models.  ANIARA studied concepts and methods for adapting and selecting ML models for mitigating ML-model performance degradation due to expected vertical and horizontal scaling in the edge infrastructure and 5G system, with specific focus on performance models for networked services, see \cite{Sanz2022, larsson2021source, Johnsson2019PerformancePI},

%
%
\paragraph{Intelligent feature selection}

A challenge for AI at the edge is related to network overhead with respect to measurements and monitoring, and feature selection for improved model performance \cite{wang2021online}. A key enabler for ML models is timely access to reliable data, in terms of features, which require pervasive measurement points throughout the network. However, excessive monitoring is associated with network overhead. Using domain knowledge provides hints to find a balance between overhead reduction and identifying future ML requirements. ANIARA implemented a method of unsupervised feature selection that uses a structured approach in incorporation of the domain knowledge acquired from domain experts or previous learning experiences \cite{taghia2022pol}.

%
%
\paragraph{Distributed learning in the edge}
Multi-domain service metric prediction is a key component in the ANIARA framework. It should enable privacy-preserving sharing of knowledge between operators, and low-overhead training of models within an operator in terms of data sharing. The approach requires in-network processing capabilities to enable federated learning. Therefore ANIARA is working on a multi-domain service metric prediction framework using federated learning, corresponding to a scenario where several services are managed by a number of operators in geographically distributed locations.


%
%
\section{Edge-cloud orchestration}
\label{section_WP4}

Slicing allows provisioning of multiple services over the same infrastructure, where virtual or physical resources are interconnected to form end-to-end logical networks, also known as slices \cite{BarCN20}. Orchestrators run resource allocation algorithms to select the most suitable set of resources to satisfy the specific needs of the clients. In the edge cloud, compute resources are located close to the users, allowing provisioning of low latency services and enabling 5G Ultra-Reliable Low-Latency Communication slices. Allocating backup resources requires protecting the slices against link or node failures. Backup resources can be provided by means of a dedicated protection scheme, where resources are dedicated for each slice. Since backup resources are accessed only in case of failures, shared protection schemes can be developed, where backup resources are shared among different slices to decrease the overall amount of required resources.  

Therefore, ANIARA has developed a heuristic-based shared protection to encourage sharing of backup connectivity and cloud resources. We evaluated this against a dedicated protection scheme using a Python simulator, published in \cite{AmaONDM21}. Results show that the shared approach reduces the blocking probability by order of magnitude, and is especially beneficial when in-node processing resources are scarce.


%
%
\section{Edge-cloud power research}
\label{section_WP5}

Installing thousands of edge data centers, primarily in cities will require significant amounts of power, however many power grids are already utilized close to 100\%. To maintain high availability, the edge-data centers need to be complemented with alternative power sources and pro-active power management systems. This requires tailored hardware solutions integrated with the power grid and on-site power generation. Supporting active load balancing by going off-grid for shorter periods of time. We are working on the design and implementation of a series of micro-edge-data center demonstrators for deployment at industrial sites. The first generation consists of a double rack configuration including, cooling, UPS-system with batteries, multiple power source inputs and IT-hardware. 

To build out a large-scale 5G edge system, smart power utilization is required. One approach is to utilize the on-site UPS installation to go off-grid during peak power periods. The battery storage needs to be dimensioned to address this active usage. Incentivizing the active participation from the edge data centers in the load-balance activity is necessary. Moving away from fixed to dynamic prices will permits battery charging during off-peaks and discharging during higher-priced periods. Discharging implies less grid power.

The value for a power grid operator will be larger than that reflected by the customer energy price, if the data center is placed in a particularly energy-hungry section of the grid. Therefore, we have initiated a dialog with a major power grid operator.

%
%
\section{Related Work}


The first incarnation of this paper appeared in \cite{Marsh2021Evolving5A} whilst other Edge roadmap papers include \cite{ding_et_al:DagRep.11.7.76, DagstuhlRoadmap2021}.  
A survey on transfer learning from methodological point of view is provided in \cite{Weiss2016ASO}, and \cite{SolorioFernndez2019ARO} gives a recent survey on transfer learning for future wireless networks. A review on techniques for unsupervised feature selection is provided in \cite{SolorioFernndez2019ARO}. A review on online feature selection techniques is given in \cite{Hu2016ASO}. The works in \cite{Lim2020FederatedLI, Xia2021ASO} give review federated learning in mobile edge networks and edge computing, respectively. Our work builds upon these works, and extends the scope specifically towards edge clouds for telecom. 

A seminal paper from four major browser companies who collaboratively designed WebAssembly can be found in \cite{Haas2017}. A recent Ph.D describes an edge computing system which offloads computations from web-supported devices to edge server \cite{HyukJin2020}. It exploits the portability of web apps when migrating the \textit{execution state} of web apps.  Edgedancer, a platform that offers infrastructure support for portable, provider-independent, and secure migration of edge services, it is a lightweight and generic execution environment by utilising WebAssembly, see \cite{Nieke2021:Edgedancer}. A thorough survey on container technologies can be found in \cite{https://doi.org/10.1002/cpe.5668}.

Resource allocation strategies for 5G networks and reliable services have been investigated recently. In particular, different techniques for backup protection of optical network resources, relying on both DP and SP schemes, have been presented in \cite{Sha20, Mar20}. Works  propose efficient DP and SP algorithms for cloud and baseband resources in 5G access and metro networks, see \cite{Kho19, Cha20}.
Considering connectivity and compute resources separately may lead to impractical solutions, especially when resources are scarce. Our work focuses on the dynamic slice provisioning where both type of resources are jointly allocated. A whitepaper presents an overview of the market and implementation trends, see \cite{Sote2021}.

%
%
\section{Future work}
\label{summary} 

Going forward in the AI for edge field, we will study distributed learning under various sources of data and system heterogeneity. The objective is to tackle concerns with data privacy, resource heterogeneity among AI actors, challenges in re-usability of previously learned ML models, and difficulties in effective incorporation of domain knowledge.  Experiments with WebAssembly based runtimes to implement \textit{code-once, execute anywhere} approaches across the device-edge-cloud continuum are ongoing.  The idea is to support offloading applications from the user equipment device to the edge node. Future work for the Feature store is a Kubernetes operator for RonDB and using it to store and serve our WASM containers. Improving RonDB and implementing an \textsl{evaluation store}, feature drift detection is planned. Before a widespread edge data center can be used, we will need to work on power integration aspects that means deployment/installation of the physical hardware. Installation at hard-to-reach places, requiring easy assembly on-site and an autonomous operation with minimal on-site maintenance is ongoing (a demo was shown at the mid-term review, April 2022). We will also investigate the potentials and limitations of resource sharing in bare metal deployments of containers, and enhanced scaling strategies to improve utilization.

%
%
\section*{Acknowledgements} 

We thank EU-CELTIC office under the project ID C2019/3-2. Also to the reviewers for feedback at mid-term review meeting, Berlin April 2022. We also acknowledge the Swedish, UK and Germany financing bodies: Vinnova AB "Automation of Network edge Infrastructure and Applications with aRtificiAl intelligence" 2020-00763, InnovateUK under the project ID 106197: ukANIARA,  and the Bundesministerium für Bildung und Forschung under the name "AI-NET ANIARA 16KIF1274K". We thank Björn Skubic at Ericsson Research for his comments on an earlier version of this paper. 
\bibliographystyle{ACM-Reference-Format}
\bibliography{sncnw2022}
\end{document}